# Multi-principal element grain boundaries: Stabilizing nanocrystalline grains with thick amorphous complexions

Charlette M. Grigorian [a], Timothy J. Rupert [a, b, c, *]

[a] Department of Chemical and Biomolecular Engineering, University of California, Irvine, California 92697, USA
[b] Department of Mechanical and Aerospace Engineering, University of California, Irvine, California 92697, USA
[c] Department of Materials Science and Engineering, University of California, Irvine, California 92697, USA
[*] Email address: trupert@uci.edu

## Abstract

Amorphous complexions have recently been demonstrated to simultaneously enhance the ductility and stability of certain nanocrystalline alloys. In this study, three quinary alloys (Cu-Zr-Hf-Mo-Nb, Cu-Zr-Hf-Nb-Ti, and Cu-Zr-Hf-Mo-W) are studied to test the hypothesis that increasing the chemical complexity of the grain boundaries will result in thicker amorphous complexions and further stabilize a nanocrystalline microstructure. Significant boundary segregation of Zr, Nb, and Ti is observed in the Cu-Zr-Hf-Nb-Ti alloy, which creates a quaternary interfacial composition that limits average grain size to 63 nm even after 1 week at ~97% of the melting temperature. This high level of thermal stability is attributed to the complex grain boundary chemistry and amorphous structure resulting from multi-component segregation. High resolution transmission electron microscopy reveals that the increased chemical complexity of the grain boundary region in the Cu-Zr-Hf-Nb-Ti alloy results in an average amorphous complexion thickness of 2.44 nm, approximately 44% and 32% thicker than amorphous complexions previously observed in Cu-Zr and Cu-Zr-Hf alloys.

**Introduction**

    The recent identification of grain boundary complexions provides new context to understand previously unexplained interfacial phenomena in materials. Grain boundary complexions are phase-like structures which can only exist in equilibrium with their abutting crystallites, fundamentally different from the traditional Gibbs definition of a bulk phase [1, 2]. Six complexion types in doped alumina samples were classified by Dillon and Harmer, listed here in order of increasing thickness: (I) sub-monolayer segregation, (II) clean grain boundaries, (III) bilayer segregation, (IV) multilayer segregation, (V) nanoscale intergranular films, and (VI) wetting films [1]. Complexion types I-IV typically have an ordered structure, while types V and VI are often amorphous. Classification schemes for grain boundary complexions are not limited to such thickness-based descriptions; other classifications include complexion composition or geometry [2]. For example, intrinsic complexions can be found in pure systems, and extrinsic complexions result from adding dopants or impurities to a material.

    The equilibrium complexion type that forms is dependent on temperature, pressure, and boundary misorientation, which can all impact interfacial structure and energy [3, 4]. The phase-like behavior of grain boundaries has been widely observed and studied in recent years [5-10], especially transitions between different complexion types that can occur analogously to bulk phase transformations [11, 12]. Several different thermodynamic models of complexion transitions have been constructed to determine the equilibrium complexion type, thickness, and composition as a function of the parameters described above [12-15]. For example, grain boundary phase diagrams have been created to describe complexion transitions and help design new microstructures that take advantage of grain boundary engineering [16, 17].

Grain boundary pre-melting transitions have been reported in metallic systems such as Fe-Si-Zn [18, 19], Fe-Nd-B [20], Fe-Mn-Cu [21], Cu-Bi [22, 23], and ceramic systems such as $Si_3N_4$ [24], and $Al_2O_3$ [25, 26]. The disordered interfaces in these materials are examples of amorphous complexions, and such features are more likely to be extrinsic in nature, requiring dopant addition [27, 28]. Structurally disordered complexions have also been connected to other interesting behaviors which affect materials processing or performance. For example, liquid-like intergranular films were found in Cu-doped $TiO_2$ and the activated sintering in this system was attributed to improved mass transport along these complexions [29]. Similar activated sintering behavior has been observed in a variety of other ceramics and refractory metals with amorphous complexions [30-36].

Applying the complexion concept to bulk nanostructured metals has the potential to simultaneously enhance thermal stability and mechanical behavior. Nanocrystalline microstructures exhibit high strength compared to traditional coarse-grained metals, but are notorious for microstructural instability [37] and poor ductility [38, 39]. Unique thermal stability behavior attributable to amorphous complexions has been reported in some systems, such as nanocrystalline Ni-W where these complexions stabilize grain size at high temperatures [40]. Mo is also reported to segregate to grain boundaries in Ni and create amorphous complexions that limit coarsening and abnormal grain growth [41]. Complexion-enhanced nanostructured metals are also reported to have high strength and ductility. For example, Zhao et al. showed that increasing the concentration of Zr allowed for the formation of crystalline/amorphous Cu/Cu-Zr interfaces and significantly increased strength [42]. Amorphous complexions can increase ductility by efficiently absorbing dislocations that have moved through the lattice, slowing crack nucleation and growth [43]. A recent computational study by Pal et al. [44] investigated the cyclic loading of nanocrystalline Cu with various amorphous complexion thicknesses. These authors reported that amorphous interfaces with thicknesses of 2 nm or greater were able to effectively accommodate the strain associated with cyclic loading, by limiting the crack formation that would normally lead to sample failure. Amorphous complexions ultimately enhance the ductility and toughness of nanocrystalline metals while retaining high strength, which was first reported in Cu-Zr alloys by Khalajhedayati et al. [45].

To date, the majority of studies concerning amorphous complexions in metallic systems examine binary alloys. Generally, advanced engineering alloys are multi-component, so understanding how complexions are formed in ternary and higher-order systems is of particular interest. Preliminary work suggests that diversifying the chemistry of the grain boundary region can form thicker and more stable complexions [46]. For example, extending a model Cu-based system from two total elements to three reduced the critical cooling rate necessary to retain amorphous complexions by roughly three orders of magnitude [47]. This enhanced stability is attributable to increased frustration against crystallization, analogous to bulk metallic glasses, which are most frequently formed with and found to be stable in multi-component alloy systems [48, 49]. High-entropy alloys, comprising multiple principal elements in near-equal proportions, are similarly more likely to form solid solutions (i.e., chemically disordered) rather than intermetallic phases (i.e., chemically ordered) during solidification [50, 51]. The multi-component chemistry that is key to retaining disorder in both of these material classes should affect amorphous complexion formation and stability too. For example, Zhou et al. [52] showed that nanocrystalline alloys with complex, "high-entropy" compositions are very thermally stable. A grain size of 32 nm was measured in a Ni-Mo-Ti-Nb-Ta alloy after annealing at 900 °C for 5 h, compared to a binary Ni-Zr alloy that coarsened to an average grain size of 189 nm after the same heat treatment.

It is important to note that the chemical composition and structure of the grain boundary regions in this multi-component alloy were not directly characterized [52], so these conclusions were based on indirect evidence.

Interfacial phases with multi-component compositions may also benefit mechanical properties. For example, Wu et al. reported that the presence of nano-sized metallic glass shells surrounding grains in a nanocrystalline Al-Ni-Y alloy significantly improved the damage tolerance of the material by allowing plastic strain up to 76% in a microcompression experiment [53]. These authors also demonstrated near theoretical yield strength in a Cu-Co-Ni alloy surrounded by amorphous Fe-Si-B interfaces [54]. Similarly, Yang et al. recently developed a $Ni_{43.9}Co_{22.4}Fe_{8.8}Al_{10.7}Ti_{11.7}B_{2.5}$ superlattice alloy with high strength and ductility derived from nanoscale chemically-disordered interfaces rich in Fe, Co, and B [55]. Together, these studies reflect the potential to improve properties in multicomponent alloys through engineering of the grain boundary chemistries. While complexion diagrams have been developed to predict interfacial structure and composition in binary and ternary alloys, compositionally complex alloys are not yet as well characterized to date [56].

Our hypothesis in this study is that increased chemical complexity at the grain boundaries will encourage the formation of thicker amorphous complexions and can be used to stabilize nanocrystalline microstructures. Cu-Zr-Hf base alloys were doped with different combinations of Mo, W, Nb, and Ti to observe how increased compositional complexity influences thermal stability and if these dopants segregate out of the Cu matrix. The greatest thermal stability was observed in the Cu-Zr-Hf-Nb-Ti alloy, due to co-segregation of Nb, Ti, and Zr to the grain boundaries and the formation of thick amorphous complexions. The Cu-Zr-Hf-Nb-Ti alloy was found to have an average amorphous grain boundary thickness of 2.44 nm, approximately 32% thicker than in a ternary Cu-Zr-Hf alloy and 44% thicker than those measured in a binary Cu-Zr alloy. This work shows that there is a clear positive correlation between chemical complexity and amorphous complexion thickness for a constant set of processing conditions. The thermal stability imparted by these unusually thick grain boundary complexions is demonstrated by an experiment in which a nanocrystalline microstructure is maintained after an aggressive 1 week heat treatment at 97% of the alloy melting temperature.

**Results and Discussion**

High-entropy alloys comprise multiple principal elements in near-equal proportions throughout the entire material, while the metals studied here are doped to create chemically complex environments only at the grain boundary regions. This is accomplished by adding small amounts of additional elements that are known or anticipated to segregate to the boundary during heat treatment. Schuler and Rupert [57] outlined specific alloy selection criteria for binary alloys that form amorphous complexions, including a positive enthalpy of segregation, a negative enthalpy of mixing, and a large atomic size mismatch with the bulk phase. These same guidelines were shown to also work for at least one ternary Cu-Zr-Hf alloy, in which the co-segregation of Zr and Hf dopants to grain boundaries led to amorphous complexion formation [46]. Therefore, Cu doped with 2 at.% Zr and 2 at.% Hf is used as a base alloy here, with two additional dopant elements added to the base in the amount of ~2 at.% each to investigate how different dopants influence grain boundary chemistry and structure, as well as thermal stability. Mo, Nb, Ti, and W are chosen as the additional dopant elements, due to the large atomic size mismatches with Cu of 39%, 38%, 28%, and 40%, respectively. The size discrepancy is expected to enhance segregation [58, 59] and these elements have been shown previously to concentrate at the grain boundaries in Cu [57, 60, 61].

To screen each quinary alloy for further investigation, XRD patterns were collected after ball milling for 10 h and after subsequent heat treatment of the powders (Figure 1). Three annealed samples were created from each of the alloys by first applying a 5 h heat-treatment at 500 °C to promote segregation, followed by a high-temperature anneal at 950 °C for 5 min, 1 h, or 1 week to allow for boundary pre-melting. During mechanical alloying, HfC forms in all three alloys. Carbide phases are a consequence of the stearic acid process control agent used to mitigate cold welding of the particles [62]. Metallic W and Mo is observed in the alloys containing these elements, Cu-Zr-Hf-Mo-W (Figure 1(a)) and Cu-Zr-Hf-Mo-Nb (Figure 1(b)), indicating that they did not fully mix into the solid solution during the ball milling process. Similarly, a NbZr intermetallic phase is seen in the as-milled Cu-Zr-Hf-Nb-Ti sample, but to a limited extent (~1 vol.%) so that the majority of these elements is contained in the solid solution. Longer milling times (20 to 40 h) were not found to better incorporate these elements and were not further analyzed.

The XRD volume percent and grain size of the second phases is presented in Figure 2 as a function of annealing time. Here and for the rest of the paper, we note that the term 'grain' refers to individual crystallites. No larger aggregates exist in our system and the line broadening in XRD patterns comes from the individual crystallites. The grain size measurements from TEM observations presented later are derived from bright field images, in which contrast is determined by crystallographic orientation and single crystallites can be reliably identified. During heat treatment, peaks for secondary metallic phases (Mo, W, and NbZr intermetallic) disappear. The W and Zr are likely joining the Cu-based matrix or segregating to grain boundaries (not detectable in XRD). However, Mo, Nb, and Hf in these alloys form carbide phases. The Cu-Zr-Hf-Nb-Ti alloy formed less carbide than the other two alloys, and interestingly, appears to resist coarsening of this phase. After annealing for 1 week, a small amount of Ti-rich hexagonal close packed phase also begins to precipitate in the Cu-Zr-Hf-Nb-Ti alloy.

The XRD grain sizes of the face-centered cubic Cu-rich phase (the majority or matrix phase) are plotted in Figure 3(a) as a function of annealing time. After 5 min of annealing at 950 °C, the grain sizes of all three alloys are very similar (~40 nm) and within statistical accuracy for XRD measurements. However, after 1 h of annealing, the grain sizes diverge and begin to show the relative difference in thermal stability of the alloys; the Cu-Zr-Hf-Nb-Ti alloy has an average grain size of 43 nm, while the Cu-Zr-Hf-Mo-W and Cu-Zr-Hf-Mo-Nb alloys both coarsen significantly to average grain sizes of 95 and 71 nm, respectively. Even after 1 week (168 h), grains in the Cu-Zr-Hf-Nb-Ti sample grow to only 63 nm, while the average grain sizes of the Cu-Zr-Hf-Mo-W and Cu-Zr-Hf-Mo-Nb alloys continue to grow to 98 nm and 102 nm, respectively. The limit for crystallite size by XRD line broadening is roughly 100 nm [63], so it is plausible that the microstructure of these two alloys continues to coarsen beyond this resolution limit during the 1 week of annealing and are no longer nanocrystalline. A recent review of doped nanocrystalline Cu showed that no alloys reported in the literature maintained nanocrystalline microstructures beyond 70% of the melting point [64]. Therefore, the current investigation demonstrates that Cu-Zr-Hf-Nb-Ti is especially stable for this class of materials, even remaining nanocrystalline after long times at temperatures approaching the melting point. In this way, complexion-enhanced alloys show a clear advantage over other nanocrystalline alloys.

The relatively low thermal stability of the other two alloys in this study may be attributed to the more prominent precipitation of second phases during annealing. Figure 2(b) shows that the overall volume of second phases in the Cu-Zr-Hf-Mo-Nb alloy increases from roughly 5.4% in the as-milled state to 8.9% after only 1 h of annealing. For the Cu-Zr-Hf-Mo-W alloy, the increase in second phase is even greater, growing from 6% in the as-milled state to a maximum of 10.8% after 1 h of annealing. The second phases rich in alloying elements effectively reduce the dopant available at grain boundaries, which can result in coarsening. A similar effect has been previously reported in other alloy systems. For example, in a Cu-Hf alloy, 4 at.% Hf was shown to form amorphous complexions, but increasing the Hf content to 11 at.% Hf leads to carbide formation and grain coarsening. Similar behavior was also observed in nanocrystalline W-Cr alloys by Donaldson et al., where the precipitation of a Cr-rich phase reduced the amount of Cr available at grain boundaries and led to an unstable grain structure [65]. The observations of our study suggest that forming thick amorphous complexions is more complicated than simply increasing the concentration of dopant atoms in the alloy. The effects of phase separation on microstructure degradation and coarsening has been discussed by Murdoch and Schuh [66]. Their investigation led to the development of nanostructure stability maps, which help avoid second phase precipitation and grain growth during design of alloys with microstructures thermodynamically stabilized by dopant segregation [67].

The Cu-Zr-Hf-Nb-Ti sample was selected for further investigation via transmission electron microscopy (TEM) to elucidate the mechanisms that contribute to its superior grain size stability. The sample which was annealed for 5 min at 950 °C was specifically selected because it is expected to have the largest grain boundary volume and minimal second phase formation. A selected area electron diffraction (SAED) pattern and bright field TEM image of the sample region from which the pattern was taken are shown in Figures 3(b) and 3(c), respectively. The SAED pattern is consistent with the face-centered cubic, NbC, and HfC phases previously detected with XRD, and the TEM image confirms the nanocrystalline grain size of the face-centered cubic matrix phase. Figure 3(d) presents the grain size distribution measured in this sample for the face-centered cubic phase and shows an average grain size of 38 nm, which is in reasonable agreement with the measured XRD average grain size of 43 nm. Higher magnification images of the grain structure of this alloy are shown in Figure 4.

High-angle annular dark field scanning transmission electron microscopy (HAADF STEM) highlights regions where high Z elements, such as Nb, Zr, and/or Hf, segregate preferentially. The yellow arrows in Figure 5 indicate examples of this dopant segregation, with the majority of grain boundaries showing brighter contrast. Energy dispersive X-ray spectroscopy (EDS) line scans confirm the high concentrations of dopants at these boundaries in Figure 6. The yellow lines in Figures 6(a) and 6(c) indicate where data was collected for Figures 6(b) and (d). We note that the widths of the dopant peaks in Figures 6(b) and (d) are not necessarily the same as the boundary width, because the electron beam used for EDS has a larger interaction volume. While both EDS scans show dopant-rich boundaries, the exact composition does not match exactly. In Figure 6(b), segregation of Nb, Ti, and Zr is observed at the probed grain boundary. These elements form a quaternary composition with the base Cu as a fourth element. The line scan in Figure 6(d) shows no obvious Zr segregation and the grain boundary is comprised of Nb, Ti, and Cu. 50 similar measurements in this sample confirm these two dominant options for grain boundary chemistry. We hypothesize that the absence of Hf in the grain boundary is attributable to the formation of a HfC phase, which depletes the potential dopant from the matrix.

High resolution TEM micrographs of amorphous complexions in the Cu-Zr-Hf-Nb-Ti sample are shown in Figure 7. Yellow dashed lines denote amorphous-crystalline interfaces (see, e.g., [68]) where the disordered complexion meets the crystalline region on either sides. These features are not observed at every single grain boundary, as some grain boundaries have a typical ordered structure. Transitions from ordered to disordered complexion types are known to occur at specific grain boundaries, such as those boundaries with high levels of dopant segregation [4] and large crystallographic misorientation [9, 11].

Thickness measurements on 46 amorphous complexions are reported in Figure 8 as both a cumulative distribution function (Figure 8a) and a histogram (Figure 8b). The current data concerning multi-principal element grain boundaries are compared to previous reports [46] from binary (Cu-Zr) and ternary (Cu-Zr-Hf) alloys in Figure 8(a). These previous studies targeted slightly lower dopant levels than the current work (5 at.% versus 8 at.% total dopant concentration), but with the Hf precipitating out as a second phase (2% dopant level) and not contributing to grain boundary decoration, the dopant levels available at the interfaces are likely comparable. All complexions were measured through a consistent procedure using the narrowest part of each boundary in an edge-on viewing condition. The mean thickness of amorphous complexions in this sample was 2.44 nm, with a standard deviation of 1.36 nm. The quinary alloy is the most chemically complex and generates the thickest amorphous complexions reported to date, 44% larger than the binary system (mean thickness of 1.70 nm) and 32% larger than the ternary (mean thickness of 1.85 nm). Additional elements segregating to the grain boundary clearly leads to thicker amorphous complexions. Notably, there are a larger number of very thick amorphous complexions in the quinary alloy. For example, ~30% of the amorphous complexions measured in the Cu-Zr-Hf-Nb-Ti sample have a thickness greater than 3 nm, compared to only ~12% being above this threshold in the binary and ternary alloys.

The additional elements in multi-component complexions can interact by competing for sites at the grain boundary or repelling each other, phenomena known based on prior observations in traditional materials like alloy steels. For example, Mo and P co-segregate to grain boundaries in these materials, where the concentration of P at grain boundaries is closely correlated with the concentration of Mo [69]. Conversely, the addition of C in Fe-P alloys decreases the amount of P at grain boundaries, and can suppress the associated grain boundary embrittlement [70-72]. Adding Cr to this system restores this grain boundary segregation, but only in the presence of C as there appears to be no effect in Fe-P alloys [70]. Attractive or repulsive forces between dopant atoms themselves may encourage co-segregation of elements or cause the depletion of others through site competition. Prior work by Xing and coauthors reported on such dopant interactions and their influence on grain boundary segregation in model alloys [73, 74]. Co-segregation was observed in Pt-Au-Pd, even when the dopant elements were predicted to desegregate from grain boundaries in simpler binary alloys. These authors presented a model to predict whether there would be dopant depletion, site competition, or co-segregation of the dopants to grain boundaries.

Complications associated with grain boundary segregation and complexion transitions have also been investigated in high entropy alloys [75-79]. Segregation to the grain boundaries and surfaces of a $Co_{20}Ni_{20}Cr_{20}Fe_{20}Mn_{20}$ was modeled by Wynnblatt and Chatain [75]. These authors showed that Cr preferred grain boundary sites more than the other elements and that this preference diminished at higher annealing temperatures. Other studies have proposed spinodal decomposition as a mechanism to enrich some boundaries with co-segregated Mn/Ni and others with Cr [77]. Another study of a high-entropy CrMnFeCoNi alloy showed boundaries rich in Ni and Mn after annealing for 500 days at 700 °C. However, Cr-rich boundaries were absent, and the chemistry appeared to be balanced by Cr-rich precipitates [80]. These results may not be in conflict, as it is plausible that spinodal decomposition progressed to an advanced state of demixing could inspire precipitation of the excess Cr in domains far from the target chemistry.

**Conclusions**

In this study, three quinary nanocrystalline alloys were formed via mechanical alloying and annealed to probe the thermal stability and potential for amorphous complexion formation. The alloys began with a base of Cu-Zr-Hf and then added a range of dopants (from Mo, W, Nb, and Ti) in different combinations but with a consistent total doping level. The following conclusions can be drawn:

- Second phase formation was observed in the three alloys investigated in this study, primarily in the form of carbide phases, but also with some limited secondary metallic phases. The grain size stability of the matrix face-centered cubic phase observed in each alloy was inversely correlated with this formation, suggesting side-reactions depleted potential dopant species available to segregate and reduce excess grain boundary energy.

- The Cu-Zr-Hf-Nb-Ti alloy had the best thermal stability, retaining a grain size of 63 nm after a week of annealing at 950 °C, more than 97% of the melting temperature of the alloy. The superior thermal stability of this alloy is attributed to the segregation of Zr, Nb, and Ti to grain boundaries, which promotes the formation of thick amorphous complexions with four element compositions. These interfacial complexions can be termed *multi-principal element complexions*, and are on average 44% thicker than those in a previously studied binary Cu-Zr alloy and 32% thicker than those in a previously studied ternary Cu-Zr-Hf alloy. The increased complexity of the grain boundary composition enhances the stability of the amorphous complexions, preventing these features from reverting to ordered complexions during quenching from high temperatures.

- In addition to the enhanced stability of the multi-principal element complexions themselves, the increased complexion thickness imparts enhanced microstructural stability by limiting grain growth.

Processing challenges were identified that reduce the efficacy of dopant elements in the system. Future computational studies may help to identify alloy chemistries that are resistant to second phase formation, and new experimental procedures could be being designed to mitigate C contamination. Together, these effects reduced the availability of dopant elements to participate in grain boundary segregation, so minimizing one or both is expected to further strengthen the demonstrated effects of thick amorphous grain boundaries.

**Materials and Methods**

Cu-Zr-Hf-Mo-Nb, Cu-Zr-Hf-Nb-Ti, and Cu-Zr-Hf-Mo-W alloys were formed via mechanical alloying in a SPEX SamplePrep 8000M Mixer/Mill. Cu (Alfa Aesar, 99.99%, -170 + 440 mesh), Zr (Micron Metals, 99.7%, -50 mesh), Hf (Alfa Aesar, 99.8%, -100 mesh), Mo (Alfa Aesar, 99.95%, -100 mesh), Nb (Alfa Aesar, 99.8%, -325 mesh), Ti (Alfa Aesar, 99.5%, -325 mesh), and W (Alfa Aesar, 99.9%, -100 mesh) powders were alloyed in an Ar atmosphere with a hardened steel mixing vial and milling media. A 10:1 ball-to-powder mass ratio was used, and 1 wt.% of stearic acid was added as a process control agent to prevent excessive cold welding. Prior to milling each of the samples, sacrificial Cu powder with 1 wt.% stearic acid was milled for 2 h to coat the vial walls and milling media with a thin layer of Cu, protecting the engineered alloy from Fe contamination due to collisions with the hardened steel during milling. Confirmation of the chemical composition of the powders was performed using EDS in a Tescan GAIA3 scanning electron microscope (SEM). Measured compositions for each sample from EDS are shown in Table 1, showing that the samples are ~90 at.% Cu with the remainder being roughly evenly split between the dopant concentrations. While minor deviations of the measured compositions and the target values are found, these are within the ~1 at.% error expected of EDS measurements.

Alloy powders were sealed inside of quartz tubes under vacuum prior to annealing, in order to limit exposure to atmosphere which could cause the formation of additional phases (especially oxides) during heat treatment. The powders were first annealed at 500 °C for 5 h in order to encourage the segregation of the various dopant elements to the grain boundaries. This was followed by an additional annealing treatment at 950 °C for 5 min (0.083 h), 1 h, and 1 week (168 h) to evaluate how amorphous complexions participate in transport kinetics. We only quote the second annealing step when describing the powder microstructures in subsequent text, but remind the reader that all samples were first exposed to the lower temperature annealing treatment. The high temperature annealing step should provide the correct conditions for grain boundary premelting while also allowing for the investigation of the thermal stability of the microstructure against coarsening. The powder samples were quenched at the end of the anneal to freeze in the equilibrium grain boundary structure at high temperatures, including any amorphous complexions. The as-milled and annealed powders were characterized using a Rigaku SmartLab X-ray diffractometer with Cu K$\alpha$ radiation and a 1D D/teX Ultra 250 detector. Phase identification, grain size measurements, and volume percent determination of all phases were extracted from XRD patterns via Rietveld analysis using MAUD software [81]. The mean grain sizes of the Cu-rich, face-centered cubic phase in all as-milled alloy powders were in the range of 19-32 nm.

An electron transparent lamella of the Cu-Zr-Hf-Nb-Ti sample annealed at 950 °C for 5 min was created for TEM via the focused ion beam (FIB) lift-out method using a Tescan GAIA SEM/FIB. Excess Ga$^+$ ion beam damage of the lamella caused during sample preparation was removed with a final 5 kV polish. Detailed characterization of the microstructure of this sample was performed in the TEM, including bright field TEM, SAED, high resolution TEM of the grain boundary structure, HAADF STEM, and EDS were performed using a JEOL JEM-2800 microscope equipped with dual EDS detectors. The presence of second phases detected in the XRD pattern was confirmed with SAED. The average grain size obtained from the XRD pattern was also confirmed by measuring >160 grains in bright field TEM. Grain sizes from bright-field TEM micrographs were obtained by measuring the area of individual crystallites, and calculating the diameter of an equivalent sphere as the crystallite size. The distribution of alloying elements throughout the sample microstructure was observed using EDS. Line scans showed dopant segregation or depletion at grain boundaries where amorphous complexions are expected to form. The influence of alloy chemistry on the thickness of amorphous complexions was investigated by measuring the thickness of 46 amorphous complexions imaged with high resolution TEM in an edge-on condition. The edge-on condition of each grain boundary was confirmed by verifying that no change in amorphous complexion thickness is observed when viewing the boundary at an overfocus and underfocus of ± 6 nm.

**Acknowledgements**

This study was supported by the U.S. Department of Energy, Office of Basic Energy Sciences, Materials Science and Engineering Division under Award No. DE-SC0021224. SEM, FIB, TEM, and XRD work was performed at the UC Irvine Materials Research Institute (IMRI) using instrumentation funded in part by the National Science Foundation Center for Chemistry at the Space-Time Limit (CHE-0802913).

| Alloy Composition (at. %) | EDS Concentration (at. %) | | | | | | |
|---|---|---|---|---|---|---|---|
| | Cu | Zr | Hf | Mo | Nb | W | Ti |
| Cu-Zr-Hf-Mo-Nb | 89.8 | 2.4 | 2.2 | 3.0 | 2.6 | - | - |
| Cu-Zr-Hf-Mo-W | 91.1 | 2.8 | 2.2 | 2.3 | - | 1.6 | - |
| Cu-Zr-Hf-Nb-Ti | 90.3 | 2.2 | 2.4 | - | 2.9 | - | 2.2 |

**Table 1.** Chemical concentrations of each alloy investigated in this study, as obtained with SEM-EDS. Dopant concentrations obtained with SEM-EDS are typically subject to measurement error of ~1 at.%.

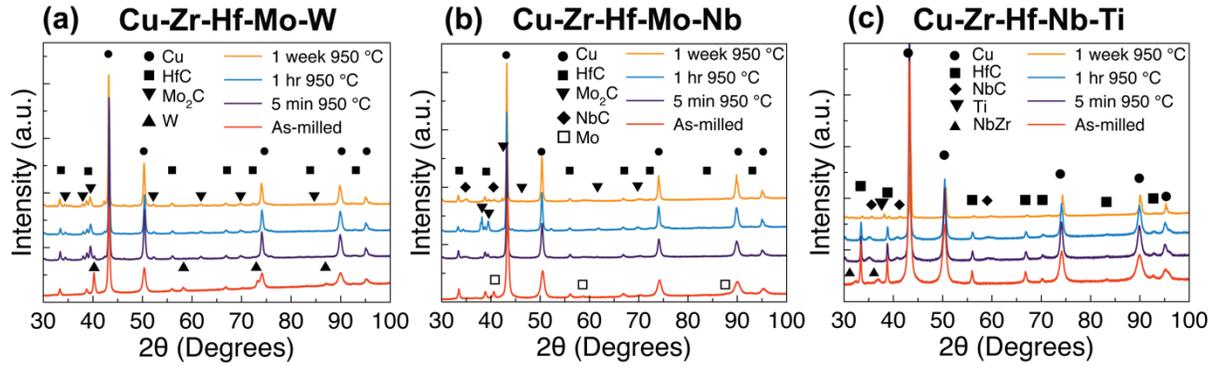

**Figure 1.** X-ray diffraction patterns of the (a) Cu-Zr-Hf-Mo-W, (b) Cu-Zr-Hf-Mo-Nb, and (c) Cu-Zr-Hf-Nb-Ti alloy samples. Patterns are shown of the as-milled powders, as well as samples subjected to an initial heat treatment at 500 °C for 5 h followed by an anneal at 950 °C for 5 min (0.083 h), 1 h, and 1 week (168 h). Phase identification shows the presence of carbide and intermetallic phases which either formed during mechanical alloying or during annealing.

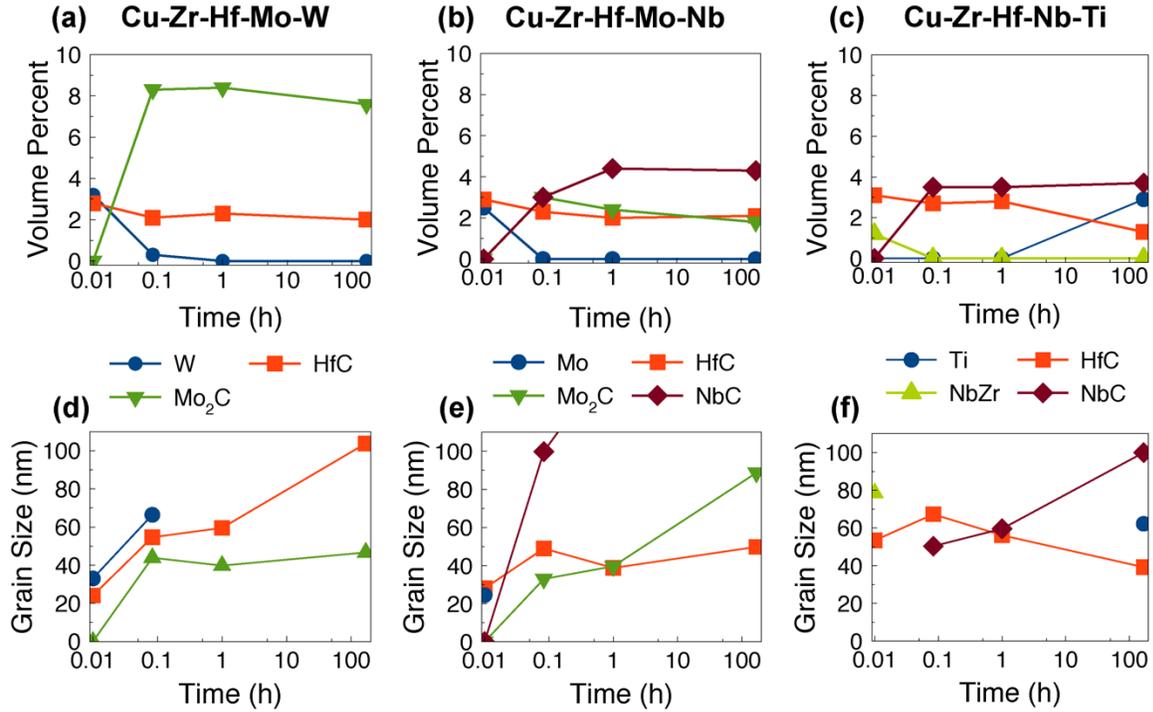

**Figure 2.** The volume percentages of second phases present in the (a) Cu-Zr-Hf-Mo-W, (b) Cu-Zr-Hf-Mo-Nb, and (c) Cu-Zr-Hf-Nb-Ti alloys plotted as a function of annealing time at 950 °C. Grain sizes of these second phases for each of the three alloys are shown in parts (d), (e), and (f), respectively. The Cu-Zr-Hf-Mo-W alloy has the greatest total volume percent of second phases, while the Cu-Zr-Hf-Nb-Ti has the lowest volume percent (<7 vol.%) and remains >93 vol.% face-centered cubic for all annealing treatments.

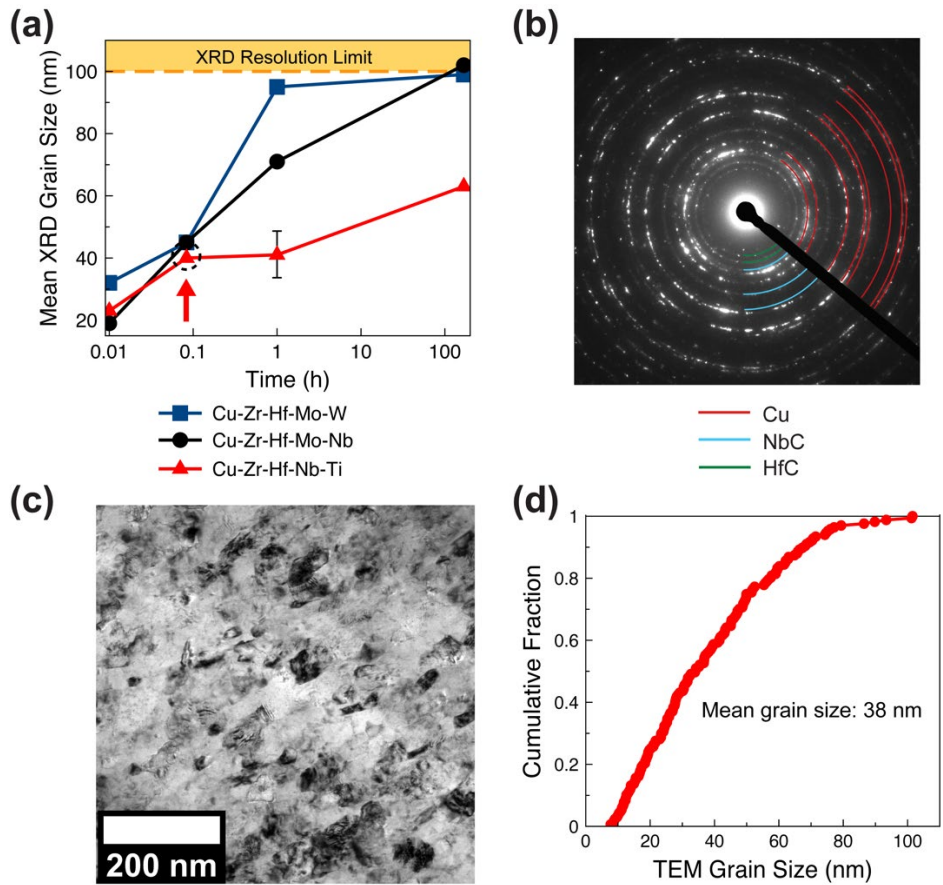

**Figure 3.** (a) Grain size of each of the three alloys plotted as a function of annealing time at 950 °C. The Cu-Zr-Hf-Nb-Ti alloy is the most stable against grain growth in this study, retaining a grain size of ~60 nm after 1 week of annealing at this high temperature. (b) SAED pattern confirming the presence of NbC and HfC phases in the Cu-Zr-Hf-Nb-Ti alloy annealed for 5 min at 950 °C. (c) Bright field TEM micrograph of the microstructure of the Cu-Zr-Hf-Nb-Ti sample annealed for 5 min. The grain size distribution measured for this sample is shown in (d), after measuring 167 grains and obtaining a mean grain size of 38 nm.

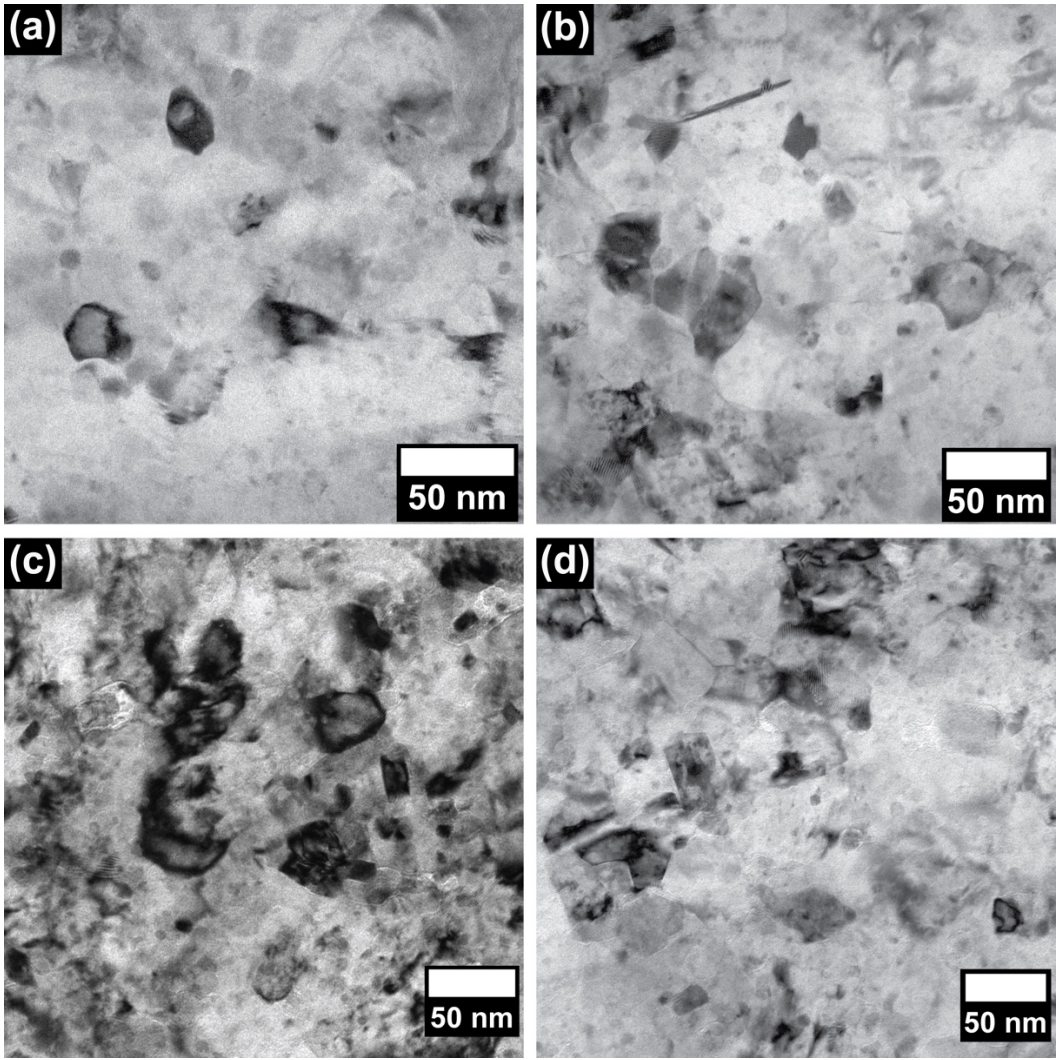

**Figure 4.** Bright field TEM micrographs of the microstructure of the Cu-Zr-Hf-Nb-Ti alloy after annealing for 5 min at 950 °C.

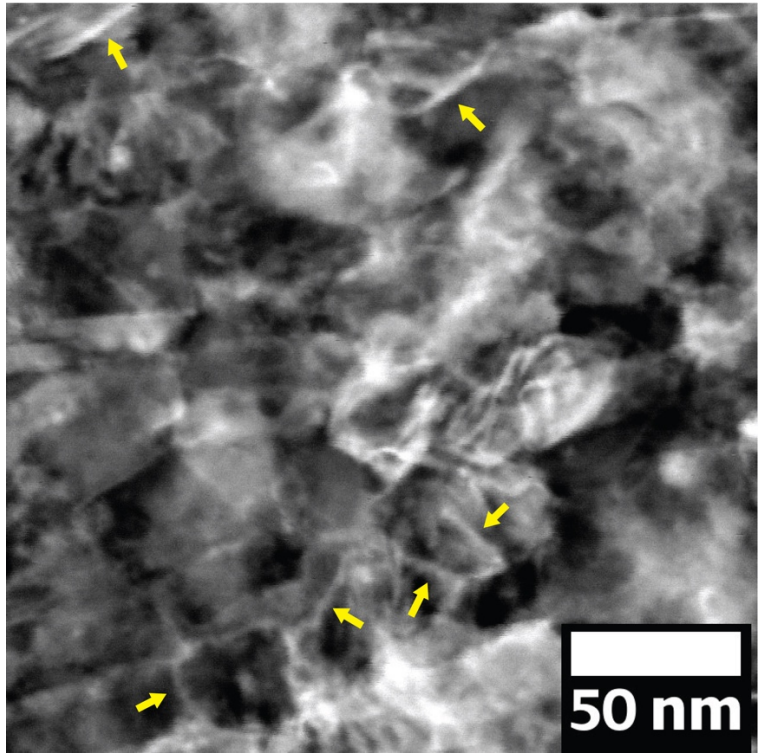

**Figure 5.** HAADF STEM micrograph of the Cu-Zr-Hf-Nb-Ti alloy. Yellow arrows point to regions of bright contrast at grain boundaries, associated with dopant segregation of higher atomic number species.

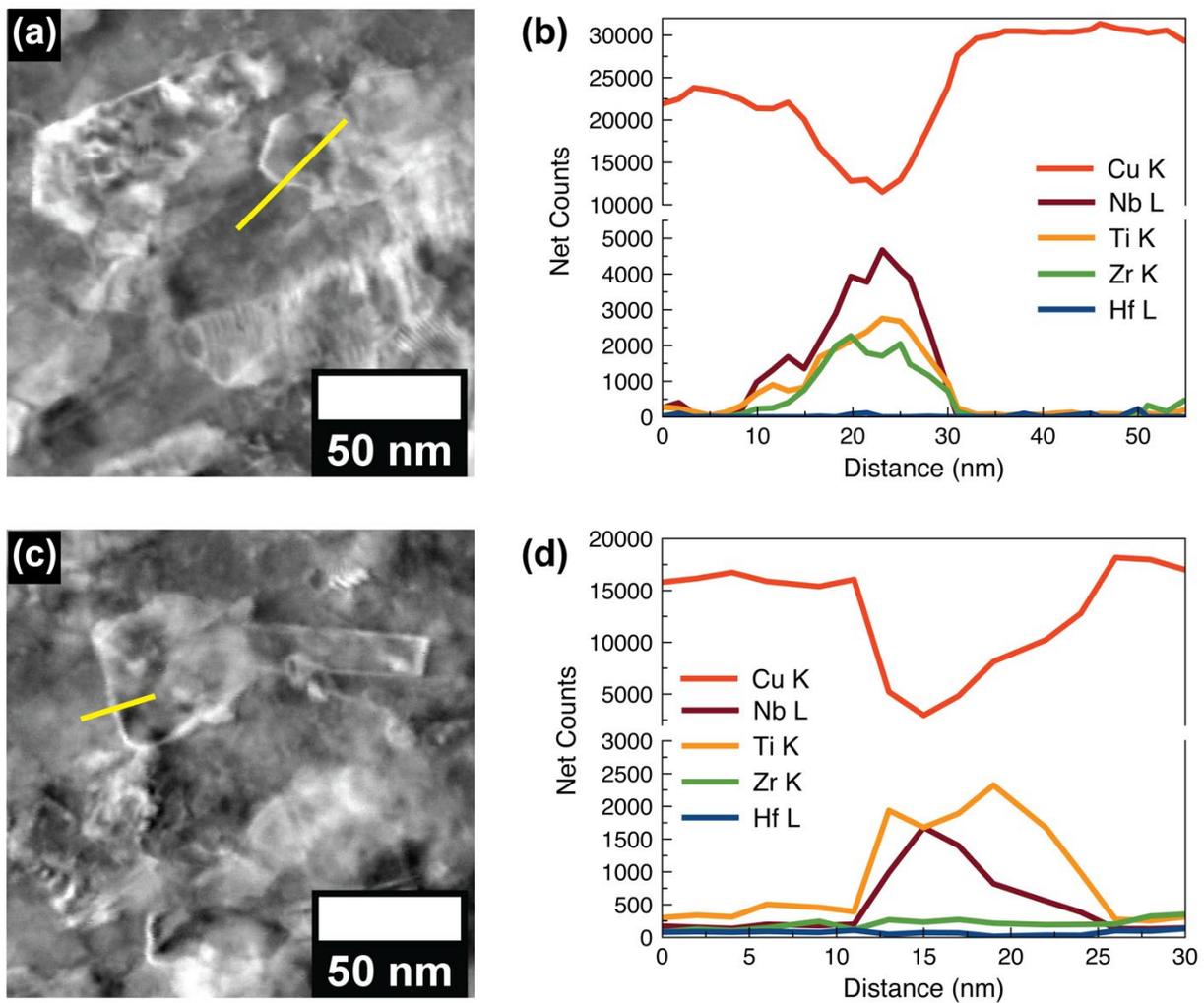

**Figure 6.** (a) HAADF STEM micrograph and (b) corresponding line scan across a grain boundary in Cu-Zr-Hf-Nb-Ti. The segregation of Nb, Ti, and Zr atoms to this grain boundary is observed for this boundary, while no Hf is measured either at the boundary or inside the grains. (b) A HAADF STEM micrograph and (c) corresponding line scan demonstrate grain boundary segregation behavior in this alloy, where the boundary is enriched with only Nb and Ti for this specific boundary.

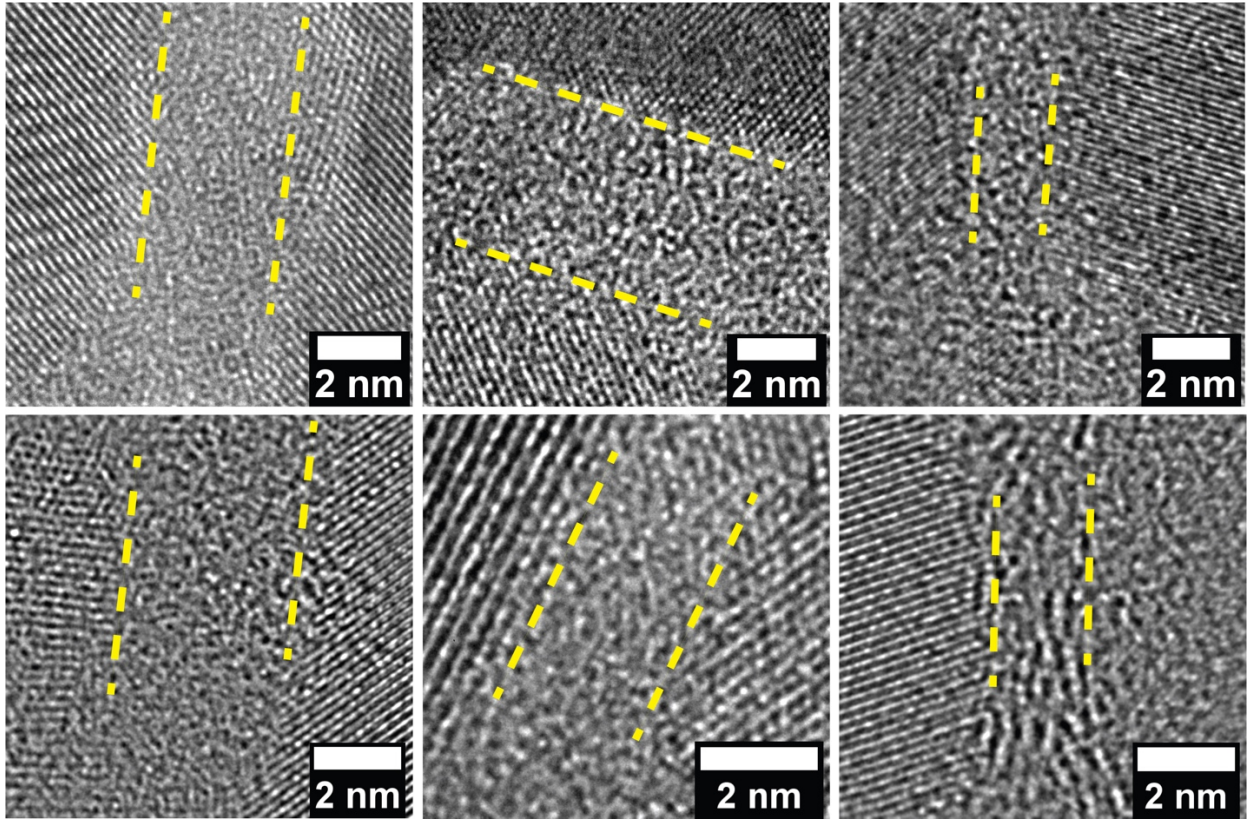

**Figure 7.** High resolution TEM micrographs of representative examples of amorphous complexions found in the Cu-Zr-Hf-Nb-Ti alloy annealed for 5 min at 950 °C. Yellow dashed lines are included to highlight where the amorphous complexion meets the crystalline regions on either side.

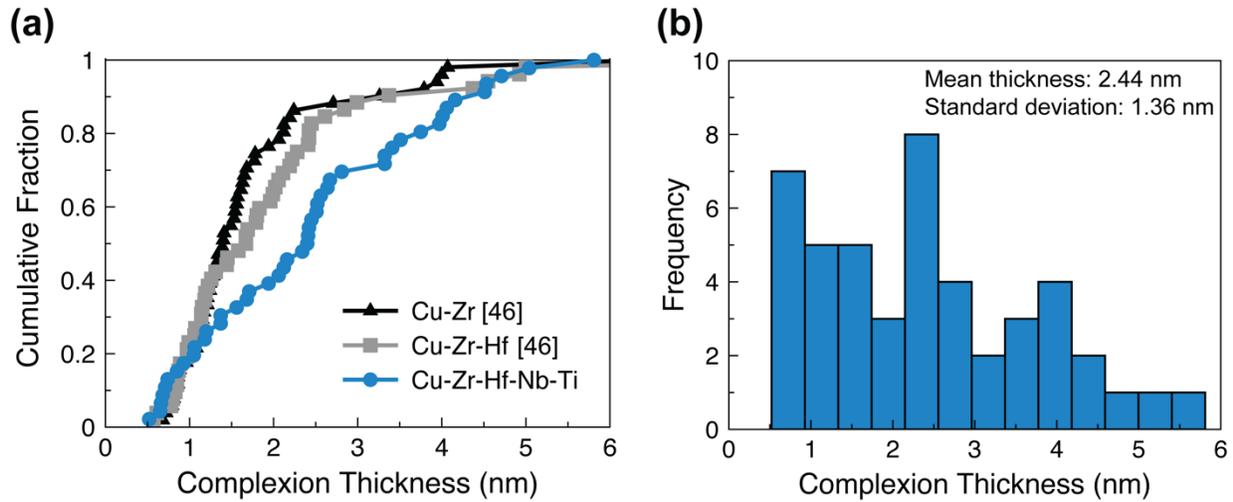

**Figure 8.** (a) Cumulative distribution function of amorphous complexion thicknesses measured from the Cu-Zr-Hf-Nb-Ti sample. 46 amorphous complexions were measured from high resolution TEM micrographs of grain boundaries in an edge-on condition. Complexion thicknesses from a prior study of Cu-Zr and Cu-Zr-Hf are plotted alongside the data from the present study to demonstrate the increased amorphous film thickness associated with a more complex grain boundary chemistry for the Cu-Zr-Hf-Nb-Ti sample. (b) Amorphous complexion thicknesses measured from the Cu-Zr-Hf-Nb-Ti sample, presented in histogram form. The mean complexion thickness is 2.44 nm, with a standard deviation of 1.36 nm.